\documentclass[aps,prc,twocolumn,showpacs,floatfix,nofootinbib,preprintnumbers,superscriptaddress,amsmath,amssymb]{revtex4-1}

\usepackage{CJKutf8}
\usepackage{epsfig}
\usepackage{color}
\usepackage{rotating}
\usepackage{natbib}
\usepackage{verbatim}
\usepackage{epstopdf}
\usepackage{graphicx}
\usepackage{changes}
\usepackage{bm}

\begin{document}

\begin{CJK*}{UTF8}{gbsn}

\title{Formation and distribution of fragments in the spontaneous fission of $^{240}$Pu}

\author{Jhilam Sadhukhan}
\affiliation{Physics Group, Variable Energy Cyclotron Centre, 1/AF Bidhan Nagar, Kolkata 700064, India}
\affiliation{Homi Bhabha National Institute, BARC Training School Complex, Anushakti Nagar, Mumbai 400094, India}
\affiliation{NSCL/FRIB Laboratory,
Michigan State University, East Lansing, Michigan 48824, USA}

\author{Chunli Zhang (张春莉)}
\affiliation{NSCL/FRIB Laboratory,
Michigan State University, East Lansing, Michigan 48824, USA}

\author{Witold Nazarewicz}
\affiliation{Department of Physics and Astronomy and FRIB Laboratory,
Michigan State University, East Lansing, Michigan 48824, USA}

\author{Nicolas Schunck}
\affiliation{Nuclear and Chemical Science Division, Lawrence Livermore National 
Laboratory, Livermore, California 94551, USA}

\date{\today}

\begin{abstract}
\begin{description}
\item[Background]
Fission is a fundamental decay mode of heavy atomic nuclei. The prevalent theoretical 
approach is based on mean-field theory and its extensions, where fission is 
modeled as a large amplitude motion of a nucleus in a multi-dimensional collective space. 
One of the important observables characterizing  fission is the charge and mass distribution 
of  fission fragments.
\item[Purpose]
The goal of this paper is to better understand the structure of  fission 
fragment distributions by investigating the competition between  the static structure of the collective manifold and  stochastic dynamics. In particular, we study the 
characteristics of the tails of yield distributions, which correspond to very 
asymmetric fission into a very heavy and a very light fragment. 
\item[Methods]
We use the stochastic Langevin framework to simulate the nuclear
evolution after the system tunnels through the multi-dimensional potential 
barrier. For a representative sample of different initial configurations along 
the outer turning-point line, we define effective fission paths by computing a large 
number of Langevin trajectories. We extract the 
relative contribution of each such path to the fragment distribution. We 
then use nucleon  localization functions along effective fission pathways to analyze the 
characteristics of prefragments at pre-scission configurations.
\item[Results]
We find that non-Newtonian Langevin trajectories, strongly impacted by the 
random force, produce the tails 
of the fission fragment distribution of $^{240}$Pu. 
The  prefragments deduced from nucleon localizations are formed early 
and change little as the nucleus evolves towards scission. On the other hand, 
the system contains many nucleons  
 that are not localized in the prefragments, even near the scission 
point. Such nucleons are rapidly distributed at scission to form the final fragments. 
Fission prefragments extracted from direct 
integration of the density and from the localization functions typically differ 
by more than 30 nucleons, even near scission. 
\item[Conclusions]
Our study shows that only theoretical models of fission that account for some 
form of dissipative/stochastic  dynamics can give an accurate description of the  
structure of  fragment distributions. In particular, it should be nearly 
impossible to predict the tails of these distributions within the standard 
formulation of time-dependent density functional theory. At the same time, the large number of non-localized 
nucleons during fission suggests that adiabatic approaches, where the interplay 
between intrinsic excitations and collective dynamics is neglected, are 
ill-suited to describe fission fragment properties, in particular their 
excitation energy.
\end{description}
\end{abstract}


\maketitle
\end{CJK*}

{\it Introduction} -- A better understanding of  nuclear fission 
 is essential for different branches of basic sciences and applications. 
Fission governs the existence and stability of  heavy and superheavy elements \cite{BLynn,(kra12),Oga15}.
In nuclear astrophysics, fission rates and the related fission fragment 
distributions are key inputs to investigate the origin of  elements heavier than iron \cite{Martinez2007,Goriely2013,Giuliani2017}. Knowledge of fission yields
is crucial for the understanding of the production rate of antineutrinos by nuclear reactors \cite{sterile}.  On the applied side, fission
data are crucial for, e.g.,  reactor design, management of the nuclear waste, and international safeguards. Since many nuclei 
relevant to nuclear astrophysics are very short-lived and out of experimental 
reach, and measurements in specific actinide nuclei for nuclear technology 
applications can pose safety issues, a predictive theoretical model for nuclear fission  is 
needed. 

Theoretical modeling of fission is extremely challenging. 
Many successful approaches to fission follow 
the original idea of Bohr and Wheeler 
\cite{bohr1939} that fission is an extreme deformation 
process typically driven by  a few collective variables that characterize the 
deformation of the nuclear surface, see Refs.~\cite{BLynn,(kra12)} for examples. 
Currently, the most 
commonly used microscopic theoretical approach to fission, rooted in an effective description of nuclear forces among nucleons, is based on nuclear density functional 
theory~\cite{schunck2016}. Here,  spontaneous fission is typically described as a  multi-dimensional quantum tunneling process through the collective space, which 
takes the nucleus from a ground-state configuration to a very deformed one at
the classical outer turning point. This approach, which is in practice 
implemented within the semi-classical (WKB)  approximation, has been  
successful in describing spontaneous fission  half-lives  and other properties \cite{(sta13),(giu14),(rob13),Tao17}. 

Calculating the characteristics of the fission fragment themselves, especially 
their distribution in charge in mass, introduces additional difficulties since it 
becomes necessary to model explicitly the dynamic from the outer turning point 
to scission. 
Although time-dependent density functional theory  methods 
may seem the most natural to describe this latter phase of  
fission process \cite{(sim14),(sca15)}, each such calculation simulates only a 
single fission event: reconstructing entire distributions can become 
prohibitively expensive especially when pairing correlations are fully taken 
into account \cite{bulgac2016}. The situation becomes more complicated for induced fission from  excited states, where pairing is quenched and  dynamics becomes strongly dissipative and non-adiabatic \cite{Norenberg}. In this regime, stochastic transport theories have been successfully applied  to describe the energy transfer between the collective and intrinsic degrees of freedom of the
fissioning nucleus \cite{Abe86,Sierk17}. Among such theories, dynamical approaches based on the Langevin equation and its derivatives have been successful in reproducing fission dynamics, including fission yields \cite{Arimoto14,Randrup15,Denisov17,Mazurek2017,Sierk17}.

{\it Theoretical framework} --
In a previous work \cite{(sad16)}, we outlined a method to calculate spontaneous-fission 
fragment distributions by combining the  multi-dimensional minimization of the 
collective action, which produces tunneling rates, with stochastic
Langevin dynamics to track fission trajectories from the outer turning point 
down to scission. The minimization of the action requires computing an 
adiabatic potential energy and the associated collective inertia in the 
multi-dimensional collective space that characterizes fission. We demonstrated that 
this two-step WKB/Langevin approach provides good agreement with experimental SF yields of $^{240}$Pu, and that the predictions are relatively 
robust with respect to changes in the dissipation tensor, ground-state 
zero-point energy, and the scission criterion.

\begin{figure}[!htb]
\includegraphics[width=1.0\columnwidth]{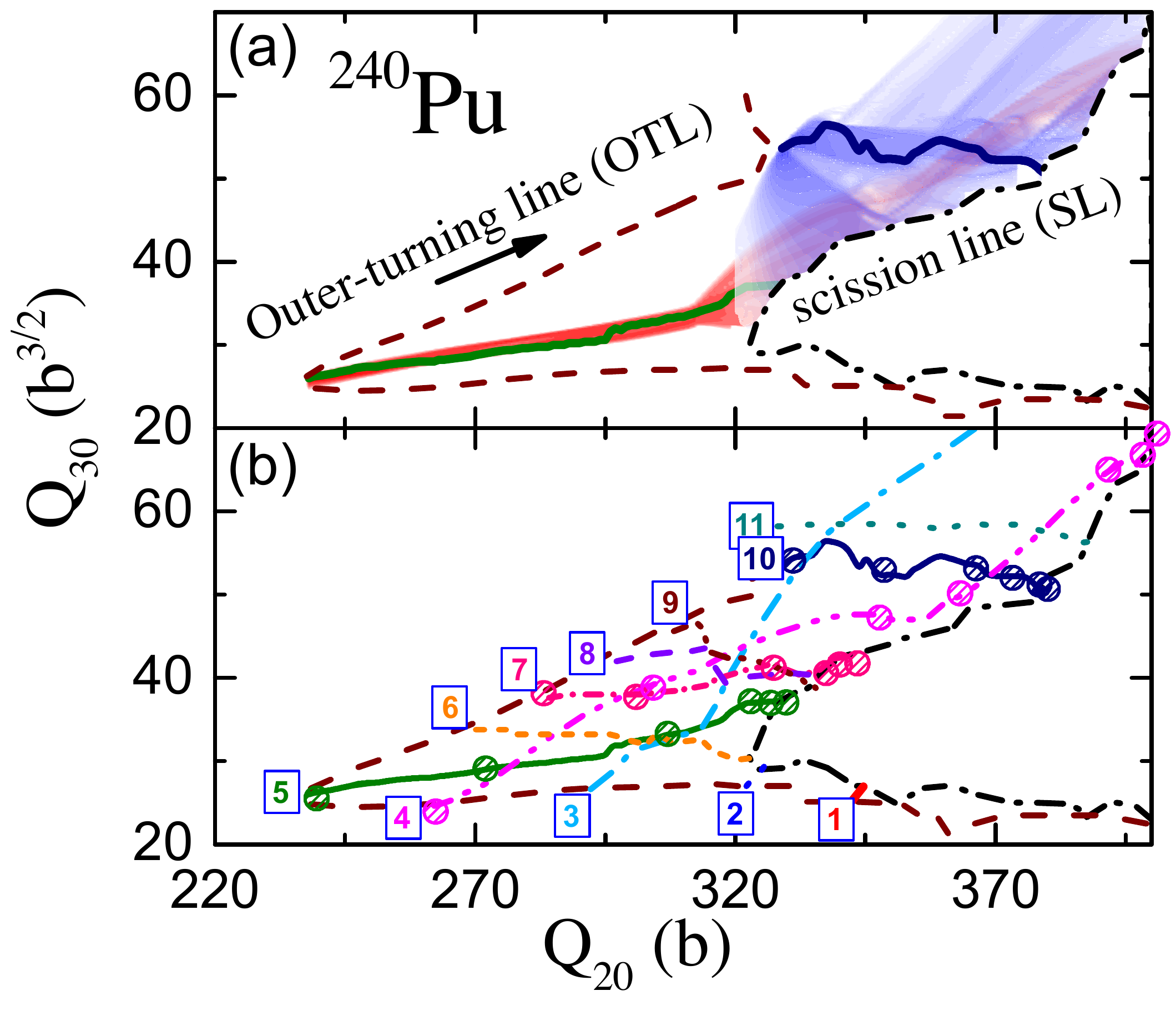}
\caption[C1]{\label{fig:EFP}
Top: the density of Langevin trajectories and the 
corresponding EFPs on the two-dimensional PES in the $(Q_{20},Q_{30})$ plane. 
The outer turning-point line (OTL) and the scission line are shown by dashed and 
dash-dotted lines, respectively. Bottom: Eleven EFPs considered in this work.}
\end{figure}

We model the collective dynamics of the 
fissioning system from the outer turning point to scission as a stochastic 
process involving the interplay between adiabatic collective motion and the 
heat bath constituted of intrinsic degrees of freedom of the nucleons. This model 
is realized effectively by solving the classical Langevin equations with both 
dissipative and random forces in addition to the standard Newtonian 
dynamics of collective variables. As a result of stochasticity, a single 
initial condition on the potential energy surface gives rise to a  
distribution of collective trajectories (because of the 
random force in the Langevin equation, each such trajectory is unique). We can define the local density of 
Langevin tracks by counting the number of trajectories in each small surface 
element $dS$ of the potential energy surface \cite{(pal08)}. We may then define an effective 
fission path  (EFP) by connecting the locus of maxima in densities of Langevin 
trajectories. The concept of EFP is illustrated in 
Fig.~\ref{fig:EFP}(a) and  eleven EFPs considered in this study are shown in 
Fig.~\ref{fig:EFP}(b). The EFP gives the most probable 
fragmentation corresponding to the associated initial configuration. The EFPs 
are calculated for different initial points on the outer turning-point line. As a 
reminder, the latter is determined from the WKB boundary condition on the 
two-dimensional potential energy surface spanned by the axial quadrupole 
($Q_{20}$) and octupole moment ($Q_{30}$). The potential energy and the 
associated non-perturbative cranking inertia tensor are calculated using the 
Skyrme energy density functional with the SkM* parametrization in the particle-hole 
channel and mixed pairing interaction in the pairing channel; see Ref.~\cite{(sad16)}  for details.

To define  prefragments formed inside the fissioning system, we compute the neutron  and proton localization functions (NLFs) $\mathcal{C}$ for the fissioning nucleus
at various pre-scission configurations along EFP.
As demonstrated in Refs.~\cite{(rei11),(zha16)}, NLFs quantify the degree of clustering more efficiently than nucleonic density distributions. This is
because  the  pattern of concentric rings exhibited by NLFs reflects the underlying shell structure, which is averaged out in density distributions.

\begin{figure}[!htb]
\includegraphics[width=1.0\columnwidth]{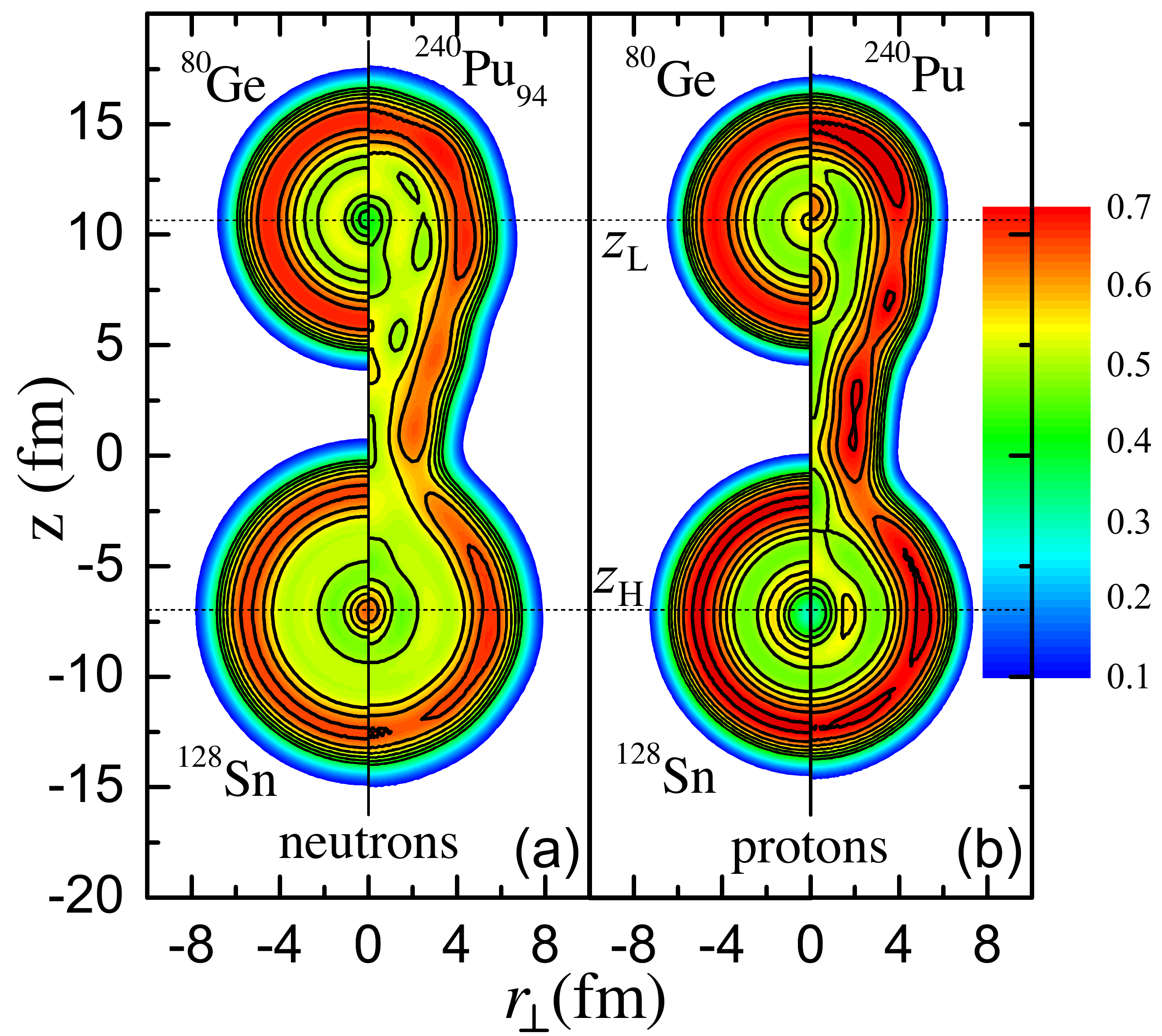}
\caption{\label{fig:NLF_example} Neutron (left) and proton (right)  localization functions
 in $^{240}$Pu at $(Q_{20}= 305$\,b, $Q_{30}=32$\,b$^{3/2})$ compared to the individual NLFs of the localized prefragments 
($^{80}$Ge and $^{128}$Sn). Symmetry axis is marked by vertical line.  Horizontal dotted lines $z_{\rm L}$ and $z_{\rm H}$  indicate the position of the  maxima  of the radial coordinate $r_\perp(z)$ associated with the NLF profile.
}
\end{figure}

Figure~\ref{fig:NLF_example} shows the typical application of NLFs to prefragment identification for a typical case of an elongate configuration of $^{240}$Pu.  The $z$-coordinates 
$z_{\rm L}$ and $z_{\rm H}$, corresponding to the maximum of the  radial coordinate $r_\perp(z)$ associated with the NLF profile, are indicated. We note that the NLFs 
for $z\geq z_{\rm L}$ and $z\leq z_{\rm H}$ exhibit characteristic ring-like pattern indicating the presence of localized nucleons inside the 
elongated $^{240}$Pu. We thus propose to define the localized prefragments by integrating the 
densities outside the horizontal lines in Fig.~\ref{fig:NLF_example} (and 
multiplying the result by two to account for reflection symmetry as the general majority of atomic 
nuclei are reflection-symmetric in their ground states). Applied to the case of Fig.~\ref{fig:NLF_example}, such a
procedure predicts that the two localized prefragments are $^{128}$Sn and $^{80}$Ge. To estimate how well 
this method works, we  perform independent calculations of  ground-state NLFs for these two prefragments. As seen of Fig.~\ref{fig:NLF_example}, there 
is a remarkable agreement between the ground-state NLFs of prefragments and 
the NLFs  of $^{240}$Pu as far as the ring pattern is concerned. It is worth noting that a fairly similar technique, based on density comparison, has been successfully applied in Refs.~\cite{(war12a),wzdeb15} to identify prefragments. 

\begin{figure}[tb]
\includegraphics[width=1.0\columnwidth]{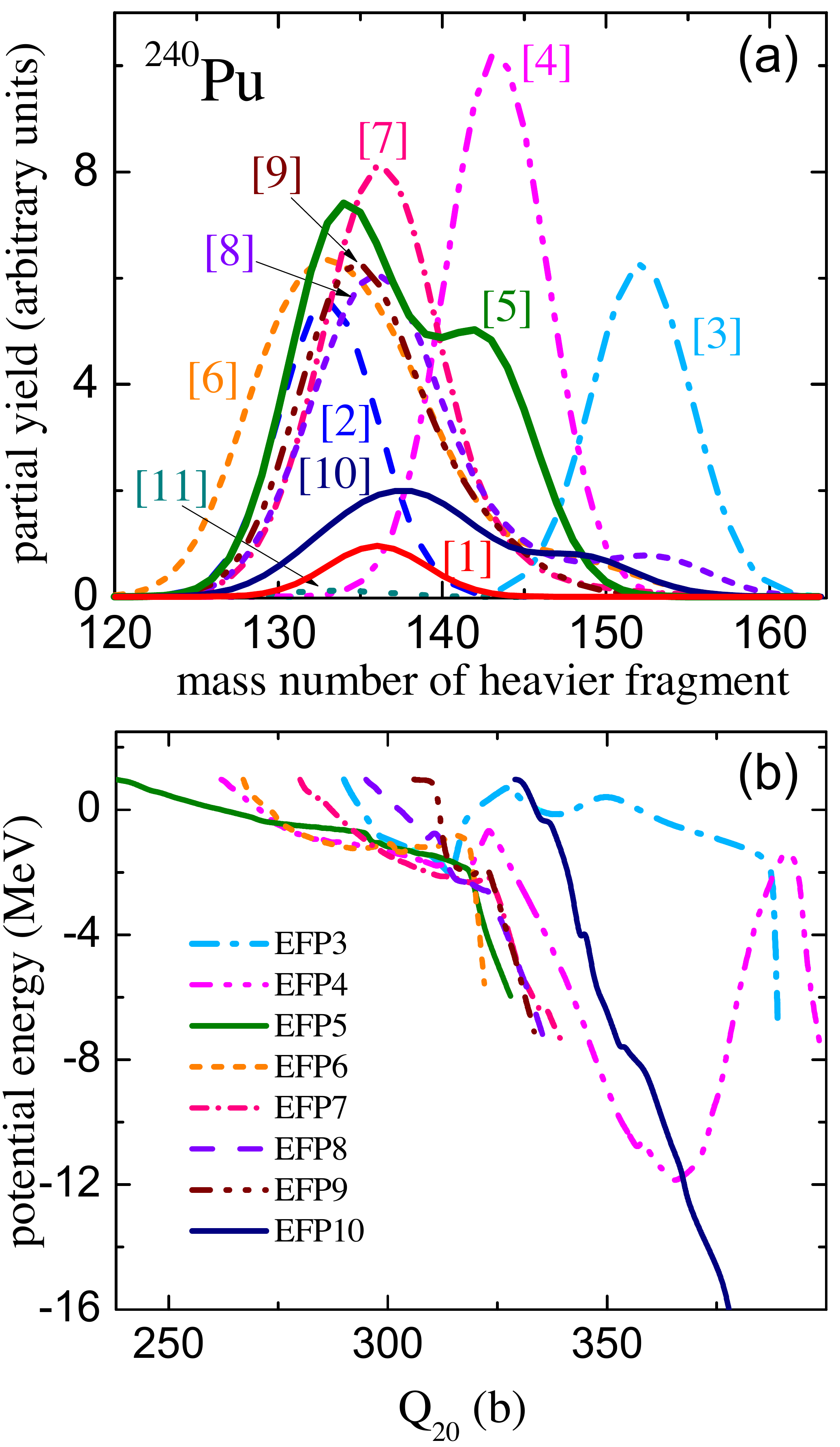}
\caption[C1]{\label{fig:FPY}
  Top: Partial mass distributions for different initial 
configurations shown in Fig.~\ref{fig:EFP}(b). The contribution  from the minimum-action fission path EFP\,5 is shown by a thick line. Bottom: 
potential energy along each EFP. Note the non-monotonic behavior for EFP\,3 and EFP\,4.}
\end{figure}

{\it Results} --  For the 11 initial configurations on the outer turning line 
shown in Fig.~\ref{fig:EFP}(b), we have generated an ensemble 
of $\approx 10^6$ Langevin tracks. We have then 
extracted the partial contribution of each EFP to the cumulative mass 
distribution by weighing the Langevin distributions with the appropriate 
tunneling probabilities determined by the collective action from the inner turning point to the outer turning point. The partial yield distributions for the heavy fragment 
mass are plotted in Fig.~\ref{fig:FPY}(a) for different initial configurations.
As expected, the peaks of the fragment mass distribution shown in Fig. 5 of 
Ref. \cite{(sad16)} are mostly built from EFPs that are near the most probable 
fission path. On the other hand, the contribution of EFPs 
that connect the region of large $Q_{30}$-values  at the outer turning line to very asymmetric 
scission configurations are quenched by  small tunneling probabilities. For 
example, the  contribution from EFP\,11 is practically negligible. 
Figure~\ref{fig:FPY}(a) shows that the tail of the yield distribution comes from EFPs that 
connect highly-probable outer turning points with very asymmetric scission 
configurations, such as those related to  EFP\,3 and EFP\,4, for which  the potential energy does not decrease monotonically, see Fig.~\ref{fig:FPY}(b). For these trajectories, the random force is primarily responsible for the upward motion, rather than the collective term. 

\begin{figure}[!htb]
\includegraphics[width=1.0\columnwidth]{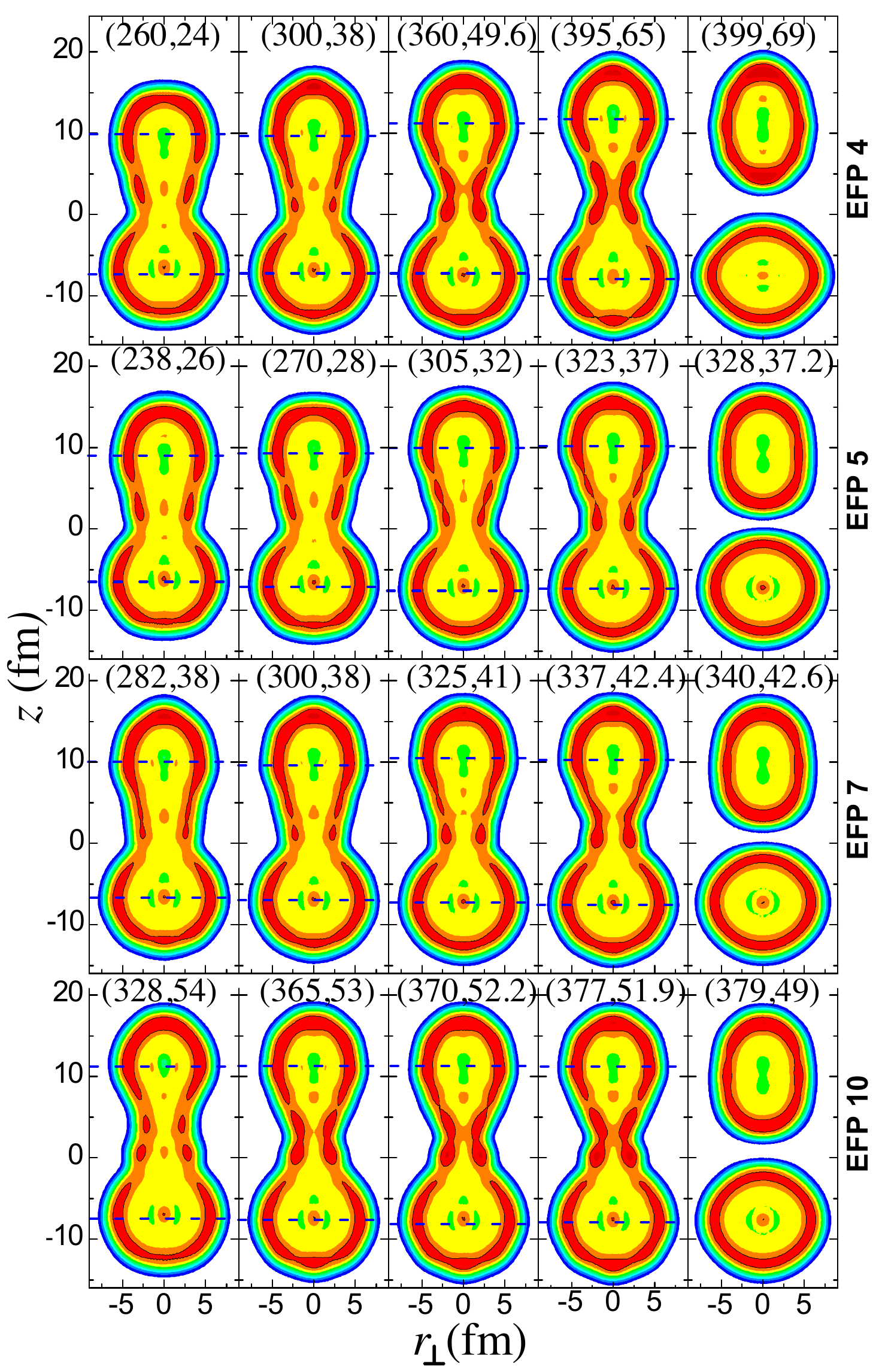}
\caption{\label{fig:NLFs}
Neutron NLFs for neutrons for several configurations (indicated by $Q_{20}$ (in b)
and $Q_{30}$ (in b$^{3/2}$)) along four EFPs indicated.
The color legend is same as in Fig.~\ref{fig:NLF_example}.
Horizontal dashed lines indicate the position of the  maxima  of the radial coordinate associated with the NLF profile.}
\end{figure}
We now focus on the structural properties of prefragments. To this end, we select five different configurations along each of the EFP\,4, EFP\,5, EFP\,7 and  EFP\,10. These configurations, representative of  various types of fission structures, are marked in Fig.~\ref{fig:EFP}(b) by circles. 
Figure~\ref{fig:NLFs} shows  the corresponding neutron NLFs; proton NLFs exhibit very similar features. It is interesting to see that   the outer structures  (at $z\geq z_{\rm L}$ and $z\leq z_{\rm H}$) of the fissioning nucleus -- associated with  prefragments --  are barely affected by the increasing shape deformation of the fissioning system; it is the neck that steadily evolves.

\begin{figure}[!htb]
\includegraphics[width=1.0\columnwidth]{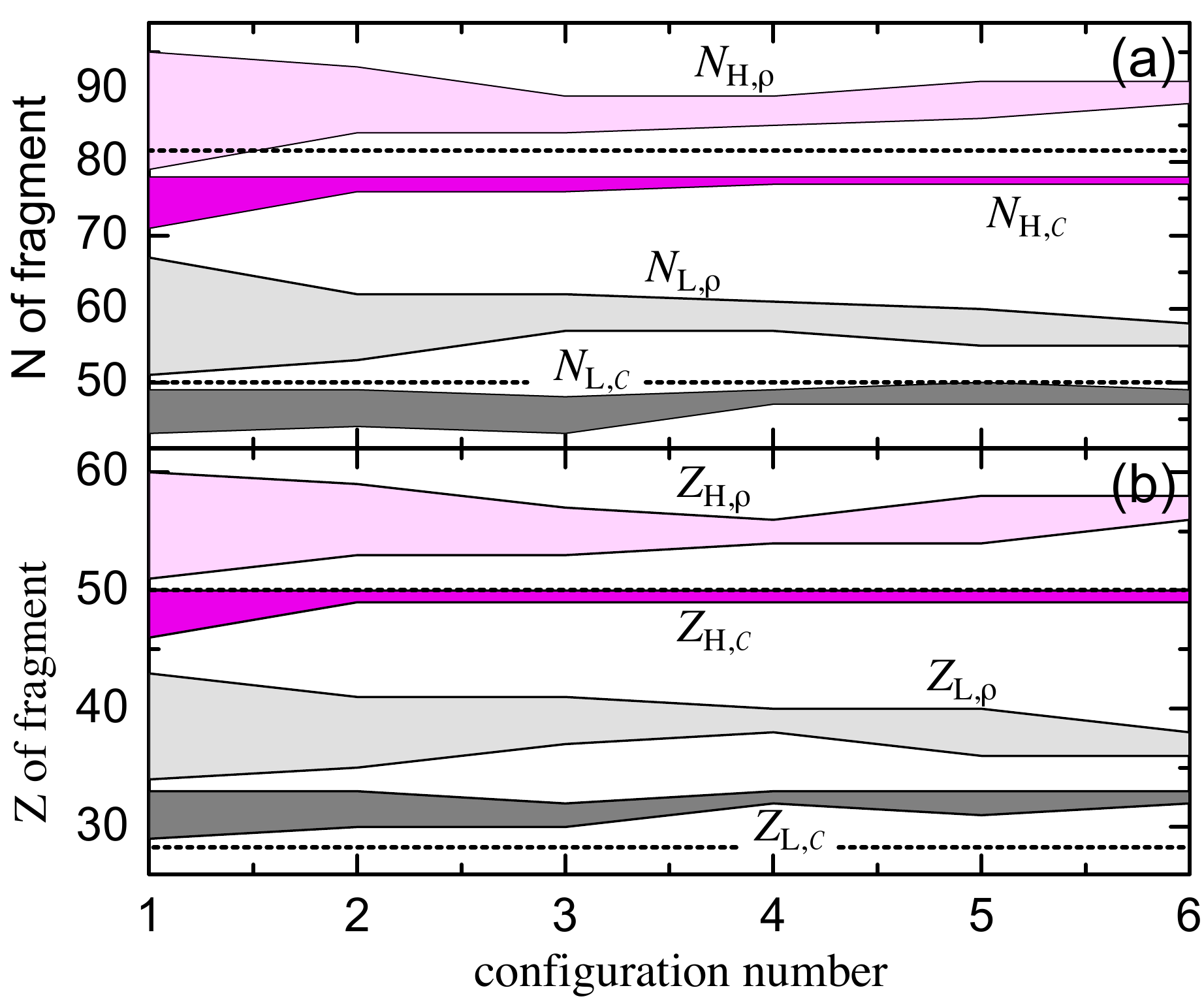}
\caption{\label{fig:partnumber}
  Upper panel: The ranges for the number of 
localized neutrons (a) and protons (b)  for heavier ($N_{{\rm H},\mathcal{C}}$, $Z_{{\rm H},\mathcal{C}}$) and lighter ($N_{{\rm L},\mathcal{C}}$, $Z_{{\rm L},\mathcal{C}}$) fragments as a 
function of the configurations along  the EFPs marked in Fig.~\ref{fig:EFP}(b) by circles. The ranges  obtained by 
integrating the nucleon density for the heavier   ($N_{{\rm H},\rho}$, $Z_{{\rm H},\rho}$) and lighter ($N_{{\rm L},\rho}$, $Z_{{\rm L},\rho}$)
fragments from the minimum of the neck. The magic numbers are marked by vertical dotted lines.}
\end{figure}
We  apply the procedure based on NLFs to extract the proton and neutron 
numbers of the  prefragments along the EFPs of  Fig.~\ref{fig:EFP}(b). The results are shown in Fig.~\ref{fig:partnumber}. We notice that the  particle numbers of the localized 
prefragments are remarkably stable as the deformation increases. In addition, 
the spread of this number, indicated by the width of the associated band,  becomes fairly narrow at large   elongations (less than $\pm 2$ particles). This  suggests that the prefragments  formed in the initial phase of
fission hardly change, and the number of nucleons, which do not belong to prefragments (primarily forming the neck) stays roughly contant in pre-scission configurations. In the case considered, around 22 neutrons and 12 protons 
are not  localized in prefragments:  they represent the glue that holds the prefragments together. For comparison, we also computed prefragments using the method of Ref.~\cite{wzdeb15} based on the analysis of linear (e.g., radially-integrated) nucleonic densities. The NLF-based approach is close to that of Ref.~\cite{wzdeb15}; in most cases, the difference does not exceed 2 nucleons. 

We have also computed the  prefragments by  integrating  the nucleonic densities  from the minimum of the 
neck  at $z=z_N$ \cite{younes2009,(sch14)}.  It is seen that the spread in the associated neutron and proton numbers is appreciably greater than for  the prefragments determined by means of NLFs, especially at smaller elongations. 
In this approach, the neck nucleons are associated with  individual prefragments; hence,   the density-based particle numbers are larger by about 5 (7) protons and 8 (11) neutrons, 
for the light (heavy) prefragments, respectively. It is to be noted that these numbers are not constant: one can see a gradual transfer of nucleons from the light prefragment to the heavy one as the scission point is approached.

\begin{figure}[htb]
\includegraphics[width=0.8\columnwidth]{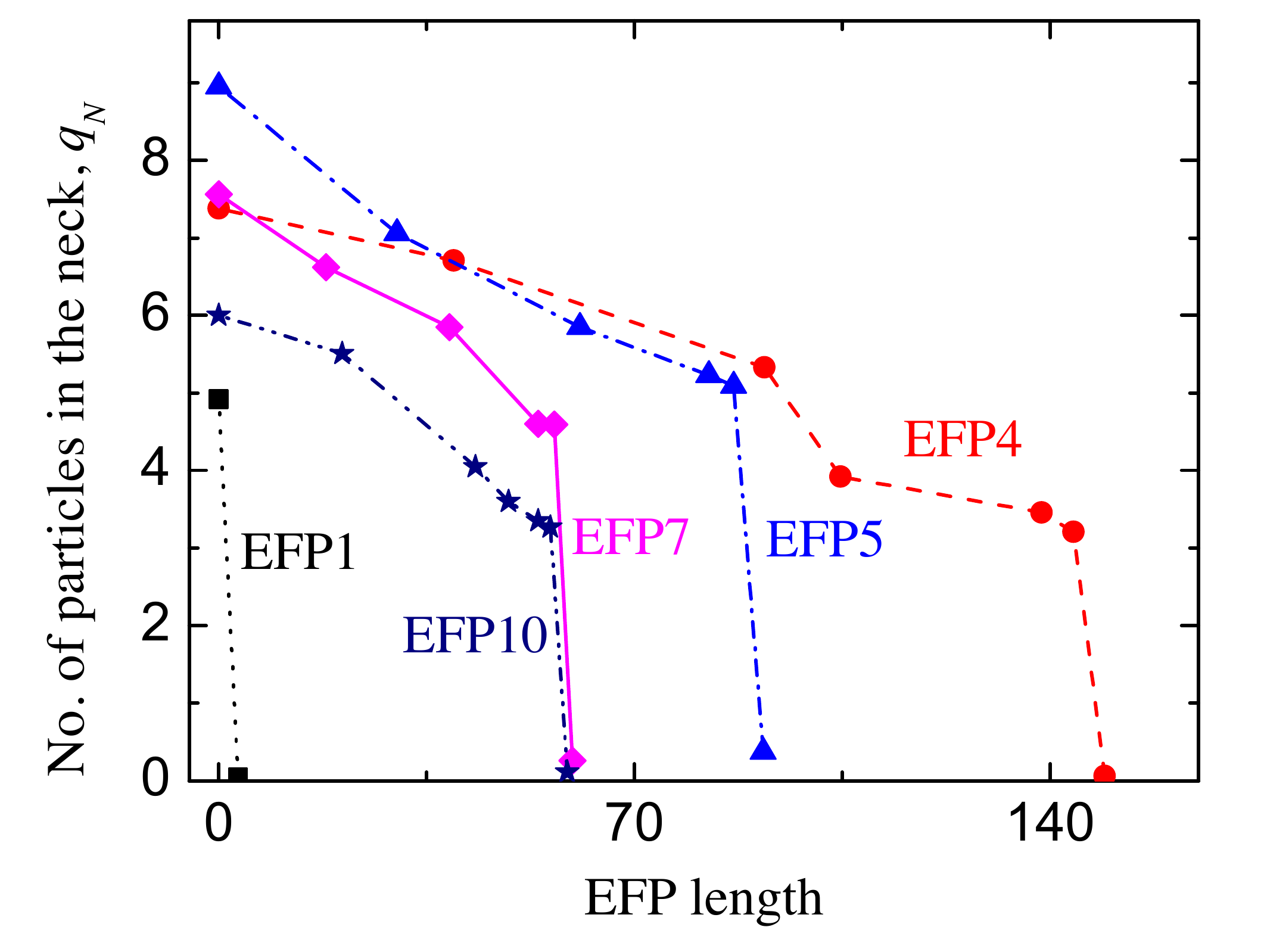}
\caption{\label{necksize}
Number of particles in the neck $q_N=\langle Q_N\rangle$   defined through the expectation value of the neck operator with $a_N$=1\,fm as a function of the length of the EFP.}
\end{figure}
Our result clearly indicates that during the motion of the system between the OTL and  scission, the distance between the early-formed prefragments gradually increases, which results in a longer and thinner neck, see Fig.~\ref{fig:NLFs}.
This neck evolution can  be illustrated by computing  the expectation value of the neck operator $Q_N=\exp[-(z-z_{\rm N})^2/a_N^2]$  \cite{Berger90,(sch14)}.
Figure~\ref{necksize} shows $q_N=\langle Q_N\rangle$ as a function of the EFP length. For all EFPs, $q_N$ gradually decreases prior to scission, at which it rapidly vanishes (see also Ref.~\cite{wzdeb15} for a similar analysis).

\begin{figure}[htb]
\includegraphics[width=1.0\columnwidth]{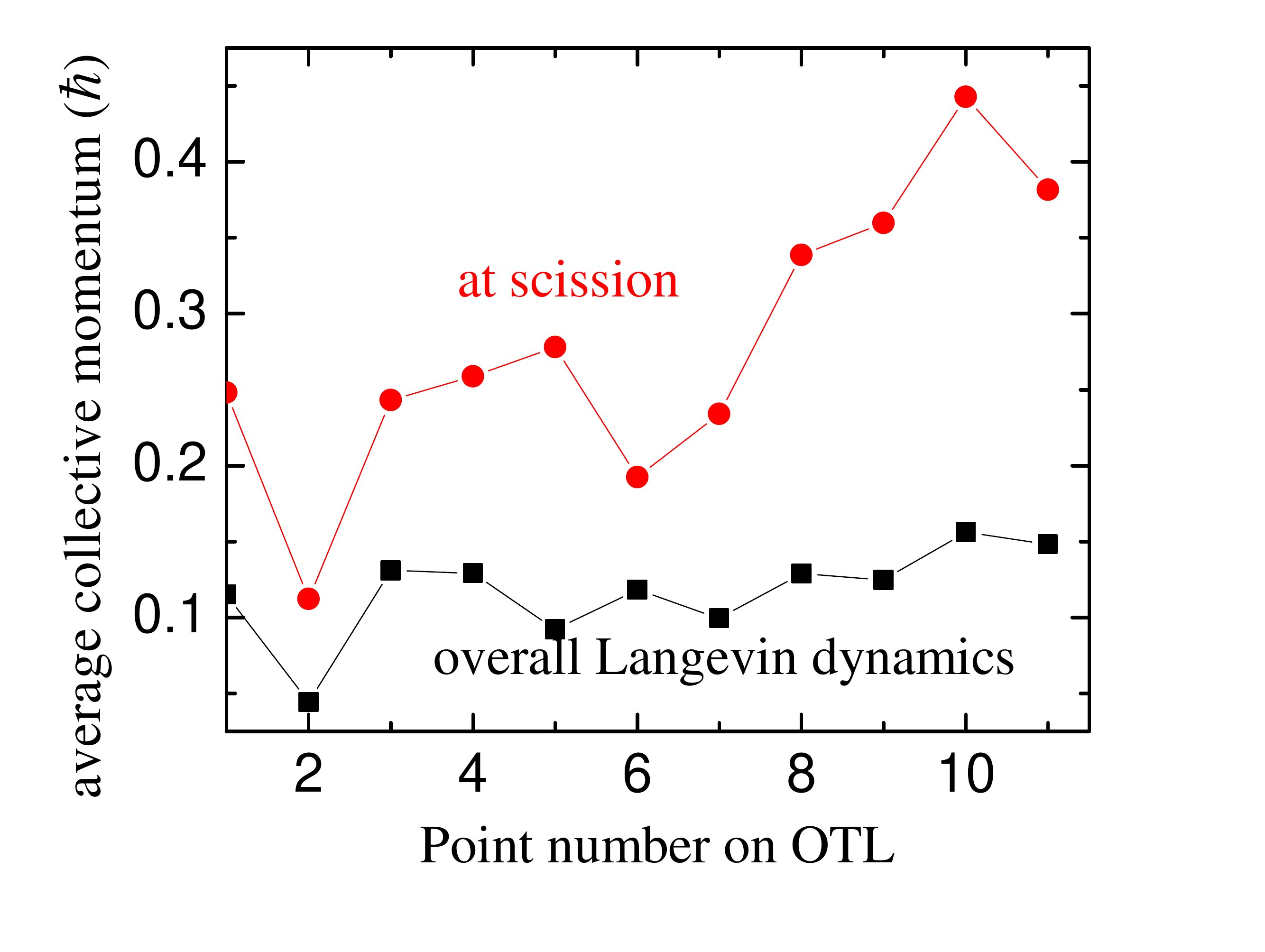}
\caption{\label{fig:collmom}
  Average collective momentum of Langevin trajectories for  different EFPs  (squares) and  shortly before reaching the scission point  (dots).}
\end{figure}
The above findings reinforce lingering questions about the process of 
fragmentation in an adiabatic theory of fission. The concept of 
scission and the criteria used to define it have long been debated in the 
literature \cite{davies1977,Brosa90,bonneau2007,(riz13)}. As our NLF-based analysis
shows, prefragments are strongly entangled near the scission 
point and a full quantum mechanical treatment is most certainly needed to 
approach the asymptotic conditions of two well-separated fragments 
\cite{(you11),(sch14)}. Recent work based on time-dependent density functional 
theory concludes that non-adiabatic collective dynamics near scission is 
essential \cite{(sim14),scamps2015,tanimura2015,bulgac2016}. To estimate the 
degree of non-adiabaticity, in Fig.~\ref{fig:collmom} we show the average collective momentum of Langevin 
trajectories for each EFP  
 together with the collective momentum prior to scission obtained by considering configurations within 3 time steps from the scission point.
It is seen that 
the average collective momentum just before scission is considerably larger than 
the average momentum of the collective coordinates outside the OTL. 

The  mechanism of generating fission yields based on the  Langevin framework proposed in Ref.~\cite{(sad16)} provides results that are consistent with experiment. In this respect, this approach seems to be superior to
the method of random neck rupture \cite{Brosa90,wzdeb15}, which  
underestimates the width of mass yields.

{\it Conclusions} --- In this work, we have performed a detailed analysis of 
the formation and distribution of spontaneous fission yields. We 
have established that the tails of the fission fragment distributions come from 
the contributions from  Langevin trajectories that 
connect highly-probable outer turning points with very asymmetric scission 
configurations. The potential energy along such  trajectories is often non-monotonic, and the corresponding  Langevin dynamics  is greatly impacted by fluctuations. This finding is consistent with the Langevin model analysis \cite{Arimoto14} that concluded that the width of the fission yields 
is primarily determined by near-scission fluctuations caused by the random force.

We have also shown that, while the  prefragments are formed early 
and change little as the nucleus deforms, there is a large number of nucleons 
($\approx 34$) that are not localized in the prefragments even near the scission 
point. These nucleons  have a large impact on the final 
properties of the fission yields as they are rapidly distributed at scission to the final fragments. The number of these nucleons depends on the neck-structure of the concerned nucleus.

Finally, our results suggest that 
even non-adiabatic time-dependent theories  will be challenged to 
reproduce the tails of the distribution unless a mechanism equivalent to the 
random force of the Langevin equation is included. Conversely, such theories 
are probably much more adapted to describing the fast rearrangements of 
nucleons at scission. It is only  very recently that a 
stochastic mean-field technique has been proposed to mimic the fluctuations within 
time-dependent  density functional theory \cite{(tani17)}.

\begin{acknowledgments} 
This work was supported by the U.S. Department of Energy under
Award Numbers DOE-DE-NA0002574 (NNSA, the Stewardship 
Science Academic Alliances program) and DE-SC0008511 (Office of Science, 
Office of Nuclear Physics  NUCLEI SciDAC-3 
collaboration). Part of this research was also performed under the auspices of 
the U.S.\ Department of Energy by Lawrence Livermore National Laboratory under 
Contract DE-AC52-07NA27344. Computational resources were provided through
an INCITE award ``Computational Nuclear Structure" by the National Center for 
Computational Sciences (NCCS), and by the National Institute for Computational 
Sciences (NICS). Computing support for this work also came from the Lawrence 
Livermore National Laboratory (LLNL) Institutional Computing Grand Challenge 
program.
\end{acknowledgments}

\bibliographystyle{apsrev4-1}
\bibliography{ref,books}

\end{document}